\documentclass[aps,amsfonts,nofootinbib]{revtex4-1}
\usepackage{epsfig}
\usepackage{graphicx}
\usepackage{slashed}
\usepackage{amsmath}
\usepackage{amsbsy}
\usepackage{bm}
\usepackage{color}
\usepackage{epsfig}

\def \be {\begin{equation}}
\def \bear {\begin{eqnarray}}
\def \eear {\end{eqnarray}}

\newcommand{\ee}{\end{equation}}
\newcommand{\bb}{\begin{equation}}
\newcommand{\eqb}{\begin{eqnarray}}
\newcommand{\eqf}{\end{eqnarray}}

\def\sigmavec{\mbox{\boldmath$\sigma$}}

\def\p{\mathbf{p}}

\newcommand{\1}{{\'{\i}}}
\def\sigmavec{\mbox{\boldmath$\sigma$}}
\def\gammavec{\mbox{\boldmath$\gamma$}}

\def\p{\mbox{\boldmath p}}

\def\nablavec{\mbox{\boldmath$\nabla$}}
\def\gammavec{\mbox{\boldmath$\gamma$}}

\def\1{\'{\i}}

\def\nablavec{\mbox{\boldmath$\nabla$}}

\def\1{\'{\i}}
\begin{document}
\title{Cosmic Four-Fermion Neutrino Secret Interactions,  Enhancement and  Total Cross Section}
\author{Dante C\'arcamo}
\affiliation{Departmento de F\1sica, Universidad de Santiago de Chile, Casilla 307, Santiago, Chile}

\author{Ashok K.  Das}
\affiliation{$^{a}$ Department of Physics and Astronomy, University of Rochester, Rochester, NY 14627-0171, USA}
\affiliation{$^{b}$ Saha Institute of Nuclear Physics, 1/AF Bidhannagar, Calcutta 700064, India}

\author{Jorge Gamboa}
\affiliation{Departmento de F\1sica, Universidad de Santiago de Chile, Casilla 307, Santiago, Chile}

\author{Fernando M\'endez}
\affiliation{Departmento de F\1sica, Universidad de Santiago de Chile, Casilla 307, Santiago, Chile}

\author{Alexios  P. Polychronakos}
\affiliation{Physics Department, City College of New York,  NY 10031, USA}

\begin{abstract} 
The scattering of neutrinos assuming a ``secret" interaction at low energy is considered.  To leading order in energy, the two-body potential is a $\delta$-potential, and it is used as a motivation to study
 generic short-range elastic interactions between neutrinos. The scattering cross section and Sommerfeld enhancement depend on two phenomenological parameters deriving from the exact form of the potential, akin to ``renormalized" coupling constants. Repulsive potentials lead to a decrease in the total cross section, resulting in an enhancement of the neutrino density. For attractive potentials of the right form, substantial Sommerfeld enhancement can appear.

\end{abstract}
\pacs{PACS numbers:}
\date{\today}
\maketitle
\section{Introduction}

In the last twenty years the nature of dark matter, needed to address the missing mass problem, has been extensively investigated \cite{lista1}. Specifically, there is a large body of evidence from astronomical observations indicating that there is more matter than can be associated with the luminous part of galaxies. This problem remains unresolved, and opens a window to new physics now called astroparticle physics \cite{lista2}. It is not yet fully understood what kind of particle dominates the dark matter and what are the mechanisms of production/annihilation for them. Neutrinos would be one possibility, but presently we know that only a small 
portion of neutrinos (1$\%$) of the CNB (cosmic neutrino background) could be  considered as dark matter. Specifically, in the CNB
the number of neutrinos per cubic centimeter is about 56 per flavor, and  this suggests that a small fraction of the dark matter could be attributable to neutrinos. Therefore, cosmic neutrinos would be responsible for only a small part of dark halos that would explain the flatness of galaxy rotation curves \cite{review1,roulet}, {\it i.e.} the constant value of the velocity of rotation of a galaxy as a function of distance rather than decreasing as expected. Even though dark neutrinos are very few,  they are nevertheless important, as their mass bounds are very robust and can be used as phenomenological data for estimating the total cosmological neutrino flux and cross sections \cite{wmap,hirano}. 

In the present state of our knowledge the cosmic neutrino background (CNB)  is a prediction of the standard cosmological model.
There are, however, known data which, when consistently interpreted, might lead to new insights beyond the standard model.
In this direction, there are several interesting questions; for example, if there are any enhancement mechanisms for dark halo neutrinos, and how they would affect physical observable processes.
 

In reference \cite{arkani} (see also \cite{iengo}) an enhancement mechanism for weakly interacting particles
was proposed, inspired by the Sommerfeld enhancement effect, and several examples illustrating how this mechanism works were
presented. A thorough treatment of the effective non-relativistic theory of dark matter long-range interactions was also presented in \cite{italia}, including their renormalization properties. In spite of the intense interest in this field in the last few years, however, to our knowledge there is no discussion on the role played by very short-range potentials (invisible at long distances), nor relevant extensions of the results presented in \cite{arkani}. 
This would be especially relevant for low energy scattering of bosons or fermions with a short-range two-body potential.
The dynamics of the CNB neutrinos at very low energy, in particular, could be considered by modeling the neutrino-neutrino interaction with a contact potential.

The goal of this paper is to provide a justification for the short range potential as stemming from possible ``secret" neutrino
interactions, to give a derivation of the scattering amplitude for such short range potentials, and to show how Sommerfeld
enhancement can emerge by computing the corresponding enhancement factor.

This work  is organized as follows. In section {\bf II} we calculate the two-body potential for neutrinos from secret interaction
in inverse powers of the neutrino mass and show that the leading term is a delta potential (contact interaction). In section {\bf III} we solve the associated scattering problem for a general short-range potential and show that physics depends on a set of
phenomenological parameters. In section {\bf IV} we consider the Sommerfeld enhancement and derive its strength in various domains of the scale of the annihilation processes. In section {\bf V} we compare and contrast our approach with standard
regularizations procedures for the delta-potential and elucidate its connection to renormalization in effective
potentials, and briefly comment on bound states. In section {\bf VI} we present our conclusions, pointing out
that the neutrino density can be substantially enhanced if the potential is repulsive, and
that there is Sommerfeld enhancement if the potential is attractive. Several calculations and justifications of approximations valid for short-range potentials are presented in the appendix.


\section{Effective two-body potential for interactions between neutrinos}
 
\begin{figure}[h]
\begin{center}
  \includegraphics[width=.4\textwidth]{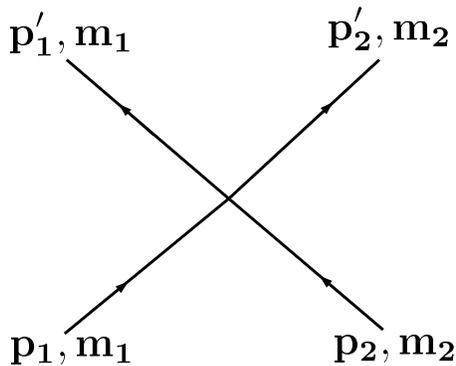}\\
  \caption{Contact interaction at the tree level.}\label{fig1}
   \end{center}
\end{figure}

Effective four-fermion interactions between neutrinos with nonstandard coupling (also called ``secret" neutrio interactions)
have been considered for quite some time \cite{lista3}, as a way to include new classes of matter as the particles mediating this interaction \cite{bere}.  Clearly the four-fermion interaction is only a first approximation, capturing the low-energy physics
of an otherwise more general interaction, valid for distances larger than some characteristic scale for the interaction. (A pure four-fermion
interaction would not even be renormalizable.) Its form is taken to be  
 \bb
- \frac{\alpha }{2M^2} ({\bar \psi}_a \gamma^\mu \psi_a)  ({\bar \psi}_b \gamma_\mu \psi_b),    \label{nolin1}
 \ee
where a sum over the three neutrino species is understood, $\alpha=\pm 1$ corresponds to an attractive/repulsive potential and  $M$ is a given mass scale.  This contact neutrino interaction leads to the Born scattering amplitude for two distinct neutrinos (see Fig. \ref{fig1})
\eqb 
M_{fi} &=& -\frac{\alpha }{M^2} {\bar u}_1 (p'_1) \gamma^\mu u_1 (p_1) {\bar u}_2 (p'_2 ) \gamma_\mu u_2 (p_2), \nonumber
\\ 
&=& \frac{\alpha }{M^2} {\bar w}_\alpha (p'_1) {\bar w}_\beta (p'_2)\, U_{\alpha \beta,\gamma \delta} \, w_\gamma (p_1) w_\delta (p_2), 
\eqf
where $w_\alpha$'s denote the positive energy spinors in the non-relativistic limit and $U_{\alpha \beta,\gamma \delta} $ is the two-body potential which will be computed in detail below. 

To extract the low-energy dynamics, we work in the non-relativistic regime $p \ll m$, with $m$ the neutrino mass. For a free massive fermion, the Foldy-Wouthyusen transformation that diagonalizes the Hamiltonian in the non-relativistic limit is given by 
\eqb 
U_{FW}&=&  \exp[{\frac{\gammavec \cdot \p}{\Large{2m}}}\theta]\nonumber\\ 
&=& \cos \left( \frac{|{\bf p}|\theta}{2m}\right) + \frac{\gammavec \cdot {\bf p}}{|{\bf p}|} \sin \left( \frac{|{\bf p}|\theta}{2m}\right),
\eqf
where 
\bb 
\theta = \frac{m}{|{\bf p}|} \tan^{-1} \left( \frac{{\bf p}}{m}\right) = 1 -  \frac{1}{3}\frac{|\bf p|^2}{m^2} + \frac{1}{5} \left( \frac{|{\bf p}|}{m^2} \right)^2 + \cdots, 
\ee
and 
\bb 
u(p) = U_{\tiny{FW}} w (p), \,\,\,\,\,\,\, w(p) = \left(\begin{array}{c} w_1(p) \\ w_2 (p) \end{array}\right). 
\ee
We use the normalization
 \[ 
  w^*_{\alpha} (p) w_{\alpha} (p)=1, 
 \]

Therefore, to order $1/m^2$ we have 
\eqb
 \cos \left( \frac{|{\bf p}|\theta}{2m}\right) &=& 1 - \frac{|{\bf p}|^2}{8m^2} + \cdots ,\nonumber\\ 
 \sin \left( \frac{|{\bf p}|\theta}{2m}\right) &=& \frac{|{\bf p}|}{2m} + \cdots,
 \eqf 
 and as a consequence 
  \bb 
  \frac{{\gammavec \cdot {\bf p}}}{{|{\bf p}|}}\sin \left( \frac{|{\bf p}|\theta}{2m}\right) \approx  \frac{\gammavec \cdot {\bf p}}{2m} + \cdots. 
    \ee
This leads to 
 \bb 
 U_{{\tiny FW}} \approx  \left(1 -  \frac{|\bf p|^2}{8m^2} \right) +  \frac{\gammavec \cdot {\bf p}}{2m} \cdots .
 \ee 

The spinor $u(p)$ can now be expanded up to order $1/m^2$ as follows
\eqb 
u(p) &=& U_{{\tiny FW}} w(p) \nonumber 
\\
&=& \left( 1- \frac{{\bf p}^2}{8m^2} \right) w +  \frac{\gammavec \cdot {\bf p}}{2m}  w + {\cal O} (1/m^3).
\eqf
Thus by restricting to the positive energy spinor space, we have to order $1/m^2$ 
\eqb 
{\bar u}_1 (p'_1) \gamma^0 u_1 (p_1) &=& u^\dagger_1 (p'_1) u_1 (p_1) \nonumber
\\ 
&=& w^*_{1\alpha} (p'_1) \left[ \delta_{\alpha \gamma} \left( 1-\frac{{\bf q}^2}{8m^2_1}\right)  + i \frac{ \left( \sigmavec \cdot \left( {\bf q} \times {\bf p}_1\right) \right)_{\alpha \gamma} 
}{4m^2_1}   \right] w_{1\gamma} (p_1), \label{am1}\eqf
where ${\bf q}= {\bf p}_1 -{\bf p}'_1=-({\bf p}_2 -{\bf p}'_2)$ is the transferred momentum. Similarly for the other spinor we have  
\eqb 
{\bar u}_2 (p'_2) \gamma^0 u_2 (p_2) &=& u^\dagger_2 (p'_2) u_2 (p_2) \nonumber
\\ 
&=& w^*_{2\beta} (p'_2) \left[ \delta_{\beta \delta } \left( 1-\frac{{\bf q}^2}{8m^2_2}\right)  - i \frac{ \left( \sigmavec \cdot \left( {\bf q} \times {\bf p}_2\right) \right)_{\beta \delta}
}{4m^2_1}   \right] w_{2\delta} (p_1) \label{am2}
\eqf
Bilinears involving $\gamma^{i}$ can also be calculated and lead to  
\eqb 
{\bar u}_1 (p'_1) \gamma^i u_1 (p_1) &=& \frac{1}{2m_1} w^*_{1\alpha } (p'_1)\left( \delta_{\alpha \beta} (-2p^i_1 + q^i) + i (\sigmavec \times {\bf q})^i_{\alpha \beta} \right) w_{1\beta} (p_1), \nonumber 
\\ 
\bar{u}_2 (p'_2) \gamma^i u_2 (p_2) &=& \frac{1}{2m_2} w^*_{2\alpha } (p'_2) \left( \delta_{\alpha \beta} (2p^i_2 + q^i) + i (\sigmavec \times {\bf q})^i_{\alpha \beta} \right) w_{2\beta} (p_2), 
\eqf
where the indices $1,2$ correspond to the two particle species and the greek indices $\alpha, \beta, \cdots$ denote spinor components.

Using the identity 
\[
{\bf p}\cdot (\sigmavec \times {\bf q}) = \sigmavec \cdot ({\bf q} \times {\bf p}), 
\]
the scattering amplitude can now be written as 
\bb 
M_{fi} = w^*_{1\alpha} (p'_1) w^{*}_{2\beta} (p'_2) U_{\alpha \beta, \gamma \delta} ({\bf q}, {\bf p}_1,{\bf p}_2)w_{1\gamma} (p_1) w_{2\delta} (p_2), 
\ee 
 where the two-body potential has the form 
 \eqb 
 U_{\alpha \beta, \gamma \delta} ({\bf q}, {\bf p}_1,{\bf p}_2)&=& - \frac{\alpha }{M^2} \bigg[ \delta_{\alpha \gamma} \delta_{\beta \delta} \left( 1-\frac{{\bf q}^2}{8} \left(\frac{1}{m^2_1}+ \frac{1}{m^2_2}\right)\right) + i \delta_{\beta \delta} 
\frac{[\sigmavec \cdot ({\bf q}\times {\bf p}_1)]_{\alpha \gamma}}{4m^2_1}  -i \delta_{\alpha \gamma} \frac{[\sigmavec \cdot ({\bf q}\times {\bf p}_2)]_{\beta \delta}}{4m^2_2} \nonumber 
\\ 
&+& \delta_{\alpha \gamma} \delta_{\beta \delta} \frac{({\bf q}^2 + 2 {\bf q}\cdot ({\bf p}_2 -{\bf p}_1) - 4 {\bf p}_1\cdot {\bf p}_2)}{4 m_1 m_2} - 
i\delta_{\beta \delta} \frac{[\sigmavec \cdot ({\bf q} \times {\bf p}_2)]_{\alpha \gamma}}{2 m_1 m_2} +
i\delta_{\alpha \gamma} \frac{[\sigmavec \cdot ({\bf q} \times {\bf p}_1)]_{\beta \delta}}{2 m_1 m_2} \nonumber 
\\ 
&+& \frac{(\sigmavec \cdot {\bf q})_{\alpha \gamma} (\sigmavec \cdot {\bf q})_{\beta \delta} 
}{4m_1 m_2} - \frac{{\bf q}^2}{4m_1m_2} (\sigmavec)_{\alpha \gamma} \cdot (\sigmavec)_{\beta \gamma} \bigg]. \label{yu1}
  \eqf
This shows that even though the particles are fermions, at leading order the dominant contribution is a contact potential given by 
\bb 
U({\bf x}) = - \frac{\alpha} {M^2} \delta ({\bf x}), \label{delta}
\ee 
where ${\bf x}$ is the relative coordinate ${\bf x}_1 -{\bf x}_2$ and the potential can be repulsive or attractive depending of the sign of $ \alpha$. Momentum-dependent higher-order corrections in this potential can  also be calculated from (\ref{yu1})
and are, generically, spin-dependent.


We can use the above contact interaction as a motivation to consider momentum-independent short-range interactions with a
range shorter than the neutrino Compton wavelength. In that regime, the full details of the interaction potential are expected
to become irrelevant and be subsumed in some macroscopic phenomenological parameters.
 
 \section{Three-Dimensional short-range potentials and their scattering properties}
 
We turn our attention to the scattering properties of short-range potentials of arbitrary form in the regime where
their range is much smaller than the de Broglie wavelength of the incident neutrinos. Such potentials would
macroscopically look like delta-functions with strength equal to their space integral. Their details, however, will in
general matter for scattering, so that more than one parameter may be relevant

We assume that the neutrinos interact with a central two-body interaction potential of the form $V(r)$ where $r$ is the distance
between the neutrinos, with the property that $V(r)$ is nonzero only within a very short range $0<r<a$. We are
interested in scattering properties for wavenumbers much smaller that $a^{-1}$.
Clearly, not all the details of
the form of $V(r)$ will be relevant. The question is: what are the relevant parameters
that fix physics for such wavelengths?

The motion in the above potential is described by the time independent Schr\"odinger equation for the
relative coordinate wavefunction
 \bb 
 \left(\nablavec^2 + k^2 \right) \psi ({\bf x}) = 2 m\, V(r)  \psi ({\bf x}). 
 \ee 
 where $k^2 =2 m E$ and $m =\frac{m_1 m_2}{m_1+m_2}$ is the reduced mass of scattering neutrinos.
The  integral scattering equation, in this case, is
 \bb
 \psi ({\bf x}) = \varphi ({\bf x}) - 2 m \,\int d^3{\bf x}' ~G ({\bf x}-{\bf x}')  ~ V(r') ~ \psi ({\bf x}'),  
\label{last1}
 \ee   
 where $\varphi ({\bf x}) = e^{i{\bf k}\cdot {\bf x}}$ represents the incident plane wave. The retarded
(outgoing) Green's function satisfies
 \bb 
 \left( \nablavec^2 + k^2 \right)  G({\bf x}-{\bf x}')  = -\delta ({\bf x} -{\bf x}')
\ee 
and is explicitly given by
\be
G({\bf x}) = \frac{1}{4\pi} \frac{ e^{ikr} }{r},
\ee
with $r = |{\bf x}|$.

In the kinematical regime of interest, only the spherically symmetric s-wave ($\ell =0$) part of the
vavefunction will matter for scattering and Sommerfeld enhancement, since higher-$\ell$ partial waves
vanish at the origin and thus are much smaller over the range of the potential than the
s-wave part. (This assertion is fully justified in the appendix.) In fact, restricting to the s-wave sector
is exact for the calculation of $\psi (0)$ since the integrand in the scattering equation in that
case is spherically symmetric for a central potential $V(r)$ and the higher angular momentum part of
$\psi ({\bf x})$ drops out. Specifically, putting ${\bf x}=0$ in (\ref{last1}) we obtain
\be
\psi(0) = 1 - \frac{m}{2\pi} \int d^3{\bf x}\,  \frac{e^{ikr}}{r} V(r) \psi ({\bf x})
=  1 - 2m \int dr\, r e^{ikr} V(r) \psi_s (r).
\label{sscat}
\ee
The s-wave part of the wavefunction, $\psi_s (r)$,  satisfies
the standard radial fixed-energy Schr\"odinger equation
\be
\psi_s (r) = \frac{\phi(r)}{r} ~,~~~
\left(\frac {d^2}{dr^2} + k^2 \right) \phi (r) = 2m V(r) \phi (r).
\ee
As usual, $\phi (r)$ vanishes at $r=0$ and $\psi (0) = \phi' (0)$. We write the full wavefunction
in terms of a normalized $u(r) = \phi (r)/\psi (0)$ as
\be
\psi ({\bf x}) = \psi(0) \frac{u(r)}{r} + \psi_{ns} (x), 
\ee
where $\psi_{ns} (x)$ is the non-s-wave part (nonzero angular momentum) of the wavefunction,
satisfying $\psi_{ns} (0) = 0$, and the function $u(r)$ is normalized to satisfy
\be
u'' + k^2 u = 2m V u ~,~~~ u(0) = 0~,~~u' (0) = 1,
\label{u}
\ee
(primes are $r$-derivatives). With these initial conditions at $r=0$ the solution $u(r)$
is unique. The coefficient $\psi(0)$ is fixed by matching the solution for $\psi({\bf x})$
to the appropriate boundary conditions at $r \to \infty$.

Plugging the above form of $\psi_s (r)$ in (\ref{sscat}) we obtain
\be
\psi(0) = 1 - 2m \psi(0) \int_0^a dr \, e^{ikr} V u, 
\ee
where we used the fact that $V(r)$ vanishes for $r>a$. This fixes $\psi(0)$ as
\be
\psi(0) = \frac{1}{1+ 2m \int_0^a dr \, e^{ikr} V u}.
\ee
Using equation (\ref{u}) for $u$ and integrating by parts we eventually obtain the exact
expression for the wavefunction at the origin
\be
\psi (0) = \frac{e^{-ika}}{u' (a) - ik u(a)}.
\label{zeroexact}
\ee

Similarly, to find the scattering amplitude we put $|{\bf x}| \gg |{\bf x'}|$ in the
scattering equation (\ref{last1}) and keep only the s-part of the wavefunction $\psi_s (r')$
to obtain
\be
\psi ({\bf x}) = \varphi ({\bf x}) - \frac{e^{ikr}}{4\pi r} \, 2m \,\int dr' d\theta' d\phi' r'^2 \sin\theta' 
 e^{-ik r' \cos \theta'} ~ V(r') ~  \psi(0)\frac{u(r')}{r'},  
 \ee
where $\theta'$ is the polar angle of $\bf x'$ measured with respect to $\bf x$. The factor
multiplying the outgoing spherical wave $e^{ikr}/r$ is the scattering amplitude $f$. Performing
the angular integral and using (\ref{u}) and integration by parts as before we obtain
\be
f = -\psi (0) \left( \frac{\sin ka}{k} \, u' (a) -\cos ka \, u(a) \right) = 
-e^{-ika} \frac{\sin ka \, u' (a) -k \cos ka \, u(a)}{k [ u' (a) - ik u(a) ]}, 
\ee
from which we obtain the scattering cross-section as
\be
\sigma = 4\pi |f|^2.
\ee

The above expressions are exact. We are interested, however, in the limit $ka \ll 1$. In this regime,
in general $k^2 \ll 2mV$ and we can treat $k^2$ as a perturbation parameter.
The full solution for $u(r)$ can then be expressed as a series in $k^2$ in a way
analogous to the Lippman-Schwinger expansion.
Specifically, we define the unique function $u_0$ satisfying:
\be
u''_0 - 2m V u_0 = 0~,~~~ u_0 (0) = 0 ~,~~ u'_0 (0) = 1, 
\ee
and the sequence of functions $u_n$, $n=1,2,\dots$
\be
u''_n - 2m V u_n = u_{n-1} ~,~~~ u_n (0) = u'_n (0) = 0.
\ee
Then the full solution for $u$ can be written as
\be
u(r) = \sum_{n=0}^\infty (ik)^{2n} u_n (r).
\ee
The above $u$ satisfies equation (\ref{u}),
and also has the correct boundary conditions at $r=0$ due to the boundary conditions
of $u_0$ and $u_n$, $n\ge 1$. Formally, the $u_n$ can be written
\be
u_n = \left(-\frac{d^2}{dr^2} + 2mV \right)^{\hskip -0.14cm -n} u_0.
\ee
The point of the above expansion is that $u_0$ represents the zero mode of the operator
$-\frac{d^2}{dr^2} + 2m V(r)$. The inverse operator
$\left(-\frac{d^2}{dr^2} + 2mV \right)^{-1}$, however, is defined on the set of states
with boundary conditions $ u (0) = u' (0) = 0$, on which this operator has no zero modes
(the unique solution with these boundary conditions is $0$), so the inverse exists.

We can now use this expansion in the expressions for $\psi(0)$ and $f$. We have
\be
\frac{1}{\psi(0)} = e^{ika} \sum_n \left[ (ik)^{2n} u'_n (a) - (ik)^{2n+1} u_n (a) \right].
\ee
The values 
\be
\lambda_{2n} = u'_n (a) ~,~~~
\lambda_{2n+1} = u_n (a),
\ee
are an (infinite) set of phenomenological parameters that
determine the scattering properties of the potential as a function of $k$.
$u_0$ behaves linearly near $r=0$ ($u_0 (r) \sim r$), while $u_n (r) \sim r^{2n+1}$
and $u'_n (r) \sim r^{2n}$.  For a {\it generic} potential such that $V a^2 \ll 1$
\be
\lambda_{2n+1} = u_n (a) \sim a^{2n+1} ~,~~~ \lambda_{2n} = u'_n (a) \sim a^{2n}.
\ee
So the leading term $\lambda_0 = 1 + O(a)$ becomes the only relevant one in the
limit $a \to 0$, the rest being negligible. Such potentials have no interesting scattering
dynamics, leading to $|\psi (0)| \simeq 1$ and $\sigma \simeq 0$.

Physically interesting potentials are these that lead to at least the first couple of the above
parameters to be of the same order of magnitude. Achieving this needs potentials whose magnitude
scales as $a^{-2}$ or faster and have nontrivial profiles.
(Examples of such potentials and the scaling of $u_n$ are given in the appendix.)
Keeping only $\lambda_0$ and $\lambda_1$
and assuming all higher $\lambda_n$  are negligible
in the limit $ka \ll 1$, we have for the wavefunction
\be
\psi(0) =  \frac{1}{\lambda_0 -ik \lambda_1 },
\label{psi0}
\ee
and the scattering amplitude and cross section
\be
f = \frac{\lambda_1}{\lambda_0 - ik \lambda_1}~,~~~
\sigma = \frac{4\pi \lambda_1^2}{\lambda_0^2 + k^2 \lambda_1^2 }.
\label{scatcross}
\ee

\section{Sommerfeld enhancement}

For short-range annihilation processes, Sommerfeld enhancement is given by
$S= |\psi (0) |^2$. In our case, however, the elastic interaction potential is also
short-range, so the above formula is not automatically true. We have to
distinguish different regimes:

a) If the range of the annihilation process $b$ is much smaller than the range
of the potential $a$, then $\psi (0)$ determines Sommerfeld enhancement:
\be
b \ll a : ~~ S = |\psi (0)|^2= \frac{1}{\lambda_0^2 + k^2 \lambda_1^2 }, 
\ee
with a maximal value for $k=0$
\be
S_{max} = \frac{1}{\lambda_0^2}.
\ee

b) If the range of the annihilation process is comparable to the range of the
potential $a$, then the values of the wavefunction within a range $a$ are relevant.
The exact formula would involve an integral of the profile of the annihilation
amplitude over the wavefunction. But as an order of magnitude:
\be
b \sim a : ~~ S\sim |\psi (a)|^2 = \left| \psi (0) \frac{u (a)}{a} \right|^2
= \frac{\lambda_1^2}{a^2 (\lambda_0^2 + k^2 \lambda_1^2 )} =
\frac{\sigma}{4\pi a^2},
\ee
with an approximate maximal value for $k=0$
\be
S_{max} \sim \frac{\lambda_1^2}{a^2 \lambda_0^2}.
\ee

c) If the range of the annihilation process is much bigger than the range of the
potential $a$, then the values of the wavefunction outside the range $a$ are relevant.
The exact formula would again depend on the exact profile of the annihilation
amplitude, but as an order of magnitude:
\be
b\gg a : ~~ S \sim \left| \frac{\int_0^b \psi (0) \frac{u(a)}{r} 4\pi r^2 dr}
{\int_0^b 4\pi r^2 dr}  \right|^2 \sim |\psi (0) |^2\frac{u(a)^2}{b^2} 
= \frac{\lambda_1^2}{b^2 (\lambda_0^2 + k^2 \lambda_1^2 )} =
\frac{\sigma}{4\pi b^2}.
\label{Slarge}
\ee
with an approximate maximum value for $k=0$
\be
S_{max} \sim \frac{\lambda_1^2}{b^2 \lambda_0^2}.
\ee
We observe that the maximun enhancement is obtained when the range of the
annihilation process is comparable to the range of the interaction.

\section{Renormalization}

The concept of renormalization appears quite often in quantum mechanics
in the context of effective interactions or singular potentials. In the first
approach, the unknown UV properties of a potential with known long-range
behavior is parametrized by adding short-range regularization terms
and fitting with known physical properties (spectra, phase shifts etc.) 
(For a nice pedagogical review see \cite{lepage}, and for recent applications
related to our considerations see \cite{italia}.)
Since the same physical properties can be obtained for various values of
the regularization terms, this leads to the idea of renormalization group.

Renormalization also appears in the context of singular potentials, such
as the delta-function potential: they are regularized with a form of cutoff
and their coupling constants are renormalized such that they give finite
physical results. Again, ``bare" constants vary with the cutoff parameter
such that they give the same physical observables.

In our approach, the issue of renormalization is, {\it a priori}, moot: we start
with a physical short-range interaction and we work out the physical
effects. There is no need to regularize anything, nor do we need to
fit any known physical data. Nevertheless, there are parallels with
renormalization that are worth pointing out.

We found that the effects of a short-range interaction can be effectively
described in terms of a set of phenomenological parameters $\lambda_n$.
For generic potentials only $\lambda_0$ matters;  for more interesting
potentials $\lambda_1$ becomes important, while for potentials with even
more nontrivial behavior higher lambdas may become relevant. This is in
the spirit of renormalization: physics is parametrized in terms of a set
of macroscopic parameters, with the full details of the short-range
interaction becoming irrelevant.

In fact, in our parameters $\lambda_0$ and $\lambda_1$ we see
analogs of both coupling constant and wavefunction renormalization:
the long-range properties of the potential, such as the 
scattering cross-section (\ref{scatcross}) and the Somerfeld
enhancement for scales much larger than the scale of the
potential (\ref{Slarge}), depend only the ratio
$\lambda_1 / \lambda_0$. This ratio can be considered as a
renormalized coupling constant, and is the only one relevant for
macroscopic physics. The wavefuction at the origin (\ref{psi0}),
however, depends on both $\lambda_0$ and $\lambda_1$, thus involving
an additional parameter. This
can be considered as a wavefunction renormalization at short ranges.

The three-dimensional $\delta$-potential and its scattering properties have
also been examined as an example of regularization and renormalization in quantum mechanics
\cite{jackiw}. The scattering equation for an exact delta function potential of
strength $g$ reads
 \eqb 
 \psi ({\bf x}) &=& \varphi ({\bf x}) - 2 mg \,\int d^3{\bf x}' ~G ({\bf x}-{\bf x}')  ~\delta ({\bf x}') ~ \psi ({\bf x}') \nonumber, \\ 
 &=& \varphi ({\bf x}) - 2 m g  ~G({\bf x}) ~\psi ({ 0}). \label{last2}
 \eqf  
  Putting ${\bf x}=0$ in (\ref{last2}) we find 
  \bb 
  \psi ({ 0}) = \frac{1}{1+2 m g\,G({0})}, \label{int}
  \ee 
  $G(0)$ is infinite, and we proceed by regularizing it. It has the momentum integral form    
  \bb 
  G(0) = \int \frac{d^3 {\bf p}}{(2 \pi)^3} \frac{1}{{\bf p}^2-k^2-i\epsilon}.   \label{green0}
  \ee 
The integral (\ref{green0}) is linearly divergent and can be regularized with a momentum cutoff
$\Lambda$ as in \cite{jackiw,thorn}, {\it i.e.} 
  \eqb 
  G(0) &=& \frac{1}{2\pi^2} \int_0^\Lambda \frac{dp\,p^2}{p^2 - k^2 -i\epsilon} \nonumber 
  \\ 
  &=&\frac{\Lambda}{2\pi^2} +\frac{i\,k}{4\pi}.\label{G}
  \eqf
Note that we would have obtained essentially the same result if we had, instead, imposed a short-distance
regularization $a$ and defined
\be
G(0)_{reg} = G(a) = \frac{e^{ika}}{4\pi a} = \frac{1}{4\pi a} + \frac{ik}{4\pi}, 
\ee
where we omitted terms of order $a$ or smaller, which identifies $\Lambda = \frac{\pi}{2a}$.
So we obtain
  \bb 
  \psi (0) = \frac{1}{1+\frac{m g\Lambda}{\pi^2}+\frac{mg}{2\pi} ik}. 
\label{psi0dir}
\ee 
From (\ref{last2}) for large $|{\bf x}|$ we identify the scattering amplitude as 
  \bb 
  f = -\frac{mg}{2\pi} \psi (0) =  \frac{1}{-\frac{2\pi}{mg}- \frac{2\Lambda}{\pi} -ik}.
\label{fdir}
  \ee

We observe that formulae (\ref{psi0dir}) and (\ref{fdir}) have the same functional form as our 
previous formulae (\ref{psi0}) and (\ref{scatcross}) upon identifying
\be
\lambda_0 = 1 + \frac{mg\Lambda}{\pi^2} ~,~~~
\lambda_1 = -\frac{mg}{2\pi}. 
\ee
Note, however, that there is {\it no} choice of scaling of the ``bare" parameter $g$
with the cutoff $\Lambda$ that can make both $\lambda_0$ and $\lambda_1$ finite.
The best one can do is demand for the long-range physical properties of the system,
such as $f$, to be finite. Thus we define
\be
\frac{2\pi}{mg}+ \frac{2\Lambda}{\pi}
= \frac{2\pi}{m{\tilde g}}, 
\label{Gtilde}
\ee
which defines a renormalized coupling constant $\tilde g$. In terms of the phenomenological
variables of last section, $\tilde g$ is essentially their ratio:
\be
{\tilde g} = -\frac{2\pi}{m}\, \frac{\lambda_1}{\lambda_0}
\ee
Written in terms of $\tilde g$, $f$ and $\sigma$ are finite:
\be
f = -\frac{m {\tilde g}}{2\pi -ikm {\tilde g}} ~,~~~
\sigma = \frac{4\pi m^2 {\tilde g}^2}{4\pi^2 + k^2 m^2 {\tilde g}^2} 
\ee 
$\psi (0)$, however, would still diverge
as $\Lambda \to \infty$ and $\tilde g$ remains finite. This calls for wavefunction renormalization
and introduces a {\it new} parameter into the problem. We define the renormalized wavefunction
at the origin
\be
{\tilde \psi (0)} = Z \psi (0) ~, ~~~ Z = -\frac{mg}{2\pi {\tilde \lambda}} = \frac{\lambda_1}{\tilde \lambda}
\label{Psitilde}
\ee
The above renormalized ${\tilde \psi} (0)$ is finite.  ${\tilde \lambda}$ 
is a new, finite parameter that plays the same role as $\lambda_1$ in the previous section. $\tilde g$ and
${\tilde \lambda}$ define the physical parameters of the system, $\tilde g$ being relevant for macroscopic
quantities and $\tilde \lambda$ for short-distance ones. The bare parameters $g$ and
$Z$ are cutoff-dependent (as manifest in (\ref{Gtilde}) and (\ref{Psitilde}) for fixed $\tilde g$
and $\tilde \lambda$) and go to zero as $\Lambda$ becomes large.

The moral of the above is that a formal renormalization procedure of a singular
delta function potential can reproduce the physics of a short-range interaction
upon proper identification of renormalized parameters. The true physics of the
situation are in the details of the short-range interaction, and its physical
content is encoded in the phenomenological parameters $\lambda_n$ that
remain relevant for wavelengths much larger than the interaction scale.

If the short-range potential is sufficiently attractive there can be bound states
(see \cite{jackiw,thorn, carlos1} for a discussion in the renormalized delta-potential case).
These are easily found through our formula (\ref{psi0}). For a bound state at $E = -B$
($B>0$) we put $ik = -\kappa$, $2mB = \kappa^2$, and look for a resonance in (\ref{psi0})
where the denominator vanishes:
\be
\lambda_0 + \kappa \lambda_1 =0 ~~ \Rightarrow~~ \kappa = -\frac{\lambda_0}{\lambda_1}. 
\ee
In the presence of a bound state, the total cross section can be expressed as 
 \bb 
    \sigma = \frac{4 \pi}{2 m B +k^2}.     \label{cross2}
    \ee 

We see that the bound state energy is determined by the same macroscopic parameter
$\lambda_1/ \lambda_0$ that determines the scattering properties (and defines the
renormalized strength in the delta-potential case). This is sensible, as a bound state
energy can be detected macroscopically without the need to probe short-distance physics
at the scale of the potential.

\section{Discussion and conclusions}

In this paper we studied the neutrino scattering problem in the low energy regime by assuming that the interaction between neutrinos is short range. As in this limit we are considering only tree level processes, the only renormalization
involved is related to the effective physics at long distances as arising from short-range properties of the potential.

We would like to emphasize  two important facts. First, for the scattering cross section, a unique ``renormalized" coupling constant
is relevant and plays a role analogous to an effective Fermi constant $G_X$. In principle, bounds for that constant 
can be found as in \cite{lista3}. The total cross section has a maximum, as can be seen from 
 
\bb
\sigma_{tot}^{\text{max}} =  4\pi G^2_X m^2_\nu, 
\ee
where $m_\nu$ represents the reduced neutrino mass. This behavior is independent of the potential being atractive or repulsive. 

Using an appropriate rescaling of the total cross section ($\sigma_{tot}/\sigma_{tot}^{\text max}$ ) and the energy ($k^2/B$),
the energy dependence of the cross section is as plotted in figure 2. 

\begin{figure}[h]
\begin{center}
  \includegraphics[width=.4\textwidth]{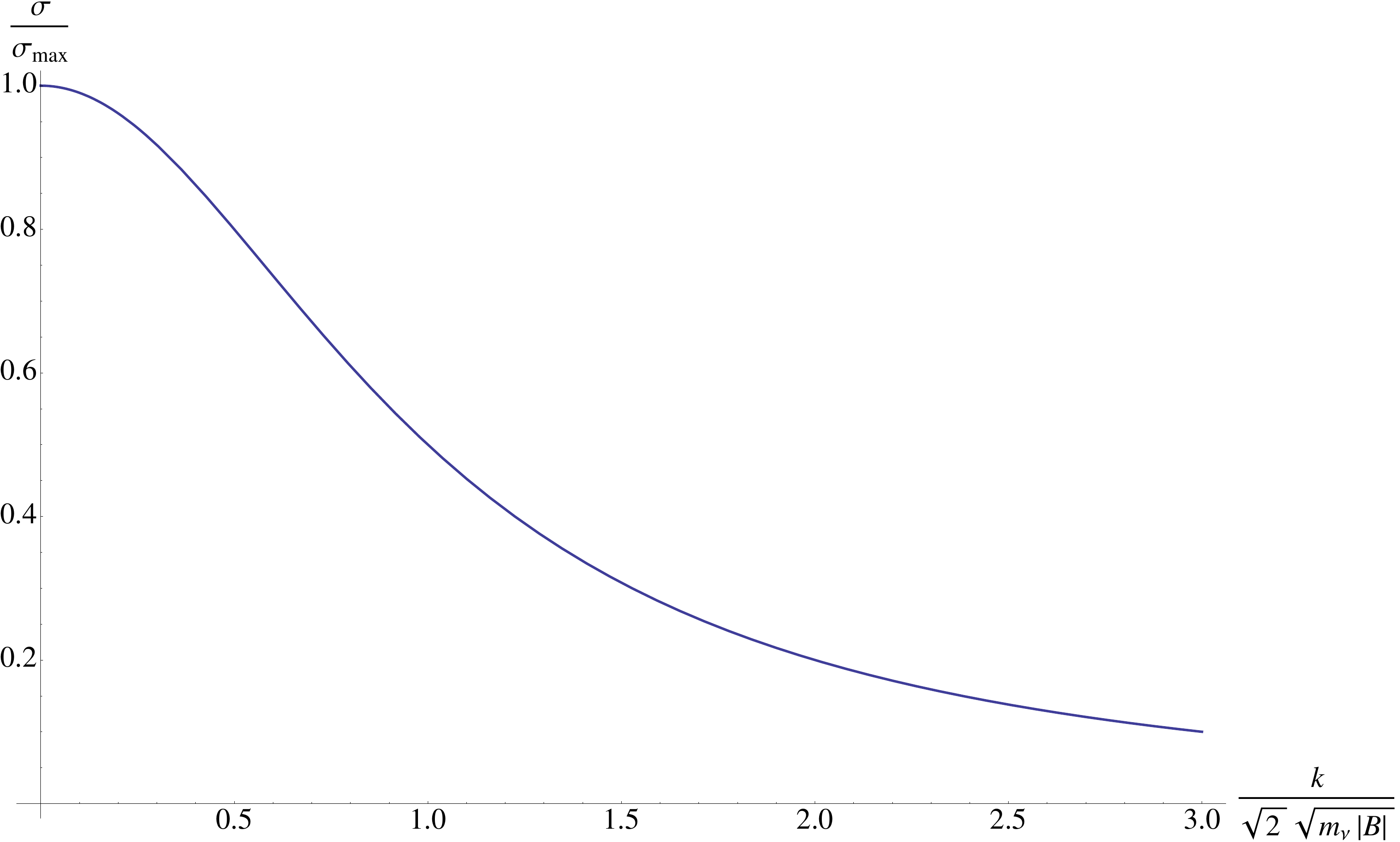}\\
  \caption{Qualitative behavior neutrino-neutrino total cross section. }\label{fig}
   \end{center}
\end{figure}




The second important point is that the Sommerfeld enhancement factor depens on additional short-range properties of the
potential, and can be quite large. So, if dark matter scattering processes occur via the mechanism proposed in \cite{arkani}, Sommerfeld enhancement could be significant. 

\acknowledgements 
We would like to thank the anonymous referee for remarks that prompted a much more
comprehensive analysis and led to substantial improvements in the content and presentation of this paper.
This work was supported by FONDECYT/Chile grants 1130020 (J.G.), 1140243  (F.M.) and  Conicyt/21140036(D. C.), by NSF grant 1213380 (A.P.) and by a PSC-CUNY grant (A.P.).

\vskip 0.2in

\centerline{\bf APPENDIX}

\vskip 0.2in

We first point our that unless the potential scales as $a^{-2}$, that is, unless $2mVa^2$ is not of negligible
magnitude, it will have no effect on scattering. We simply rescale the variable in the equation for $u(r)$
to $s = r/a$, thus expanding the range of the potential to the interval $s\in [0,1]$, and also define
\be
{\bar u} (s) = \frac{1}{a} u (as).
\ee
In terms of ${\bar u} (s)$
\be
{\bar u}'' + a^2 (k^2  - 2mV) {\bar u} = 0 ~,~~~ {\bar u}' (0) = 1.
\ee
For small $a$, the effect of the $k^2 - 2mV$ term is negligible and only the zero-order solution
${\bar u} = s$ survives, the deviation from it being of order $a^2$; that is,
\be
{\bar u} = s + a^2 \int_0^s ds' \int_0^{s'} ds'' \, [2mV(s'')- k^2 ] \, s'' + O(a^4).
\ee
So
\be
u(a) = a {\bar u}(1) = a + O(a^3) ~,~~~ u' (a) = {\bar u}' (1) = 1 + O(a^2),
\ee
leading to negligible scattering effects.

Even a generic potential scaling like $a^{-2}$ would have limited effect: ${\bar u}_0 (s)$
would be a function of order 1 and we would get
\be
u_0 (a) = a {\bar u}_0 (1) ~,~~~ u'_0 (a) = {\bar u}'_0 (1) ~ \rm{(} \neq 1 \rm{)}.
\ee
So, generically, $\lambda_1$ would still be neglibible, although $\lambda_0$ would now be
different than 1 and thus would produce a (momentum-independent) Sommerfeld enhancement
factor.

The other interesting possibility would arise if ${\bar u}'_0 (1)$ is small (so $s=1$
would be near a zero of the function $ {\bar u}'_0 (s)$) and of order $a$. In that case
both $\lambda_0$ and $\lambda_1$ are small, but their ratio is of order 1, thus
leading to nontrivial scattering effects but divergent Sommerfeld enhancement.
(This is, essentially, what the standard ``renormalization" of the delta-potential \cite{jackiw}
achieves.)

Nontrivial potentials where both $\lambda_0$ and $\lambda_1$ are finite and non-negligible
require a behavior that is stronger than $a^{-2}$ for a range of values of $r$, as
well as a change of sign in the range ($0,a$). To illustrate the possibility, we
will present a  ``proof of concept" potential, of no particular physical significance.
We start by giving the desired $u_0 (r)$, from which the appropriate potential
$2mV = u''_0 / u_0$ can be obtained. Take
\be
u_0 (r) = r + \frac{A+aB}{3} \left(\frac{r}{a}\right)^3 - \frac{A}{5} \left(\frac{r}{a}\right)^5.
\ee
(The choice of odd powers eliminates a non-essential divergence of $V$ at $r=0$.)
It is clear that the above $u_0$ satisfies $u_0 (0) = 0$ and $u'_0 (0) =1$. For $r=a$ we have
\be
u_0 (a) = \frac{2A}{15} + \frac{B+3}{3} a ~,~~~
u'_0 (a) = 1+B.
\ee
For small $a$, both $u_0 (a)$ and $u'_0 (a)$ remain nonzero and finite.
The potential corresponding to the above $u_0$ is
\be
V(r) = \frac{1}{2m}\, \frac{u''_0 (r)}{u_0 (r)} = \frac{1}{ma^2}\,
\frac{(A+aB) \left(\frac{r}{a}\right) -2 A \left(\frac{r}{a}\right)^3}
{r + \frac{A+aB}{3} \left(\frac{r}{a}\right)^3 - \frac{A}{5} \left(\frac{r}{a}\right)^5}.
\ee
The above potential in nonsingular everywhere, of order $a^{-2}$ for
most of the range $0<r<a$ but becoming of order $a^{-3}$ as $r$ nears 0. It is positive
for $r<a/\sqrt{2} +O(a^2)$ and becomes negative for $r > a/\sqrt{2}+O(a^2)$. These features are generic: a purely
positive potential cannot produce $u_0 (a)$ and $u'_0 (a)$ that are both nonzero and
finite in the limit of small $a$.

We can also show that
$u_n (a)$ and $u'_n (a)$ ($n\ge 1$) will go to zero as $a \to 0$. To start, we point out that higher
$u_n$ can be expressed recursively in terms of integrals
\be
u_n (r) = u_0 (r) \int_0^r dr' \frac{1}{u_0^2 (r' )} \int_0^{r'} dr'' u_0 (r'' ) \, u_{n-1} (r'' ).
\label{un}
\ee
The argument for the smallness of higher $u_n$ is based on the fact that
\be
u_0 (r) = f_0 (r/a), 
\ee
where $f_0 (s)$ is a function that remains finite in the limit $a \to 0$. By changing variables
$r = as$ in (\ref{un}) we see that
\be
u_1 (r) = a^2 f_1 (r/a) ~~{\rm where}~~
f_1 (s) = f_0 (s) \int_0^s ds' \frac{1}{f_0^2 (s' )} \int_0^{s'} ds'' f_0 (s'' ).
\ee
Since $f_0$ remains finite in the limit $a \to 0$, $u_1$ will scale like $a^2$
{\it unless} the integrals above develop a singularity as $a \to 0$. This, however,
is not the case: the only singularity could arise at $s' \to 0$, where $f_0 \to as'$
and the denominator diverges, but we can check that the $s'$-integral does not diverge.
Therefore, $u_1 (a) \sim a^2$.
A similar argument shows that $u'_1 (a) \sim a$, and recursively $u_n (a) \sim a^{2n}$,
$u'_n (a) \sim a^{2n-1}$, so the only survining parameters are $u_0 (a)$ and $u'_0 (a)$.
Clearly, to get potentials with nonvanishing $u_1$ or higher, we need to pick $u_0$ that
becomes divergent (of order $a^{-1}$ or higher) somewhere in the range $0<r<a$.
Such behavior is highly unphysical and provides a reason why we do not expect potentials
with nonvanishing higher $\lambda_n$ to arise.

We conclude by showing that for the scattering cross-section higher angular momentum sectors
have a negligible contribution in the limit $ka\to 0$. (For the Sommerfeld factor their contribution
is exactly zero.) The physical reason is that the radial part of the wavefunction for such sectors satisfies
\be
u'' +k^2 u =2mV u +\frac{\ell (\ell +1)}{r^2}u, 
\label{ell}
\ee
the same as for the s-wave sector but with an additional centrifugal potential. The
classical inflection point of a particle with energy $k^2/2m$ off the centrifugal potential
is at
\be
k^2 = \frac{\ell (\ell +1)}{r^2} ~~\Longrightarrow~~ r_c = \frac{\sqrt{\ell (\ell +1)}}
{k}.
\ee
Since $ka \ll 1$, $r_c\gg a$. So the region where the potential $V$ is nonzero is deep inside
the classically forbidden region of the particle. The potential is effectively shielded by the
centrifugal barrier, accessible only through tunneling effects. Only for $\ell =0$ there is
no barrier and the wavefunction can access the potential and feel its effects. In what
follows we will back this intuition with a calculation.

We will define, again, the unique solution of the radial equation (\ref{ell}) with boundary
conditions
\be
u \sim r^{\ell+1} ~{\rm as}~ r \to 0~, ~~{\rm or}~~
u^{(\ell +1 )} (0) = (\ell +1 )!~,~~ u^{(n)} (0) = 0 ~{\rm for}~ n<\ell+1, 
\ee
(exponents in parenthesis indicate derivatives). Then the radial wavefunction in this sector is
\be
\psi_\ell (r) = A \frac{u(r)}{r}, 
\label{in}
\ee
with $A$ a scale parameter that is fixed by boundary conditions at $r \to \infty$.
For $\ell =0$ (the s-wave), $A=\psi (0)$, but for higher $\ell$, $A$ does not contribute to
$\psi (0)$.

To determine the bounday conditions at $r \to \infty$ we write the plane wave decomposition
in terms of spherical harmonics and spherical Bessel functions
\be
e^{ikx} = \sum_\ell i^\ell (2\ell +1 ) j_\ell (kr) P_\ell (\theta ), 
\ee
with $\theta$ measured with respect to the axis $\vec k$. So the radial part of this plane wave
in the $u$-parametrization ($u = r e^{ikx}$) in the $\ell$ sector is
\be
u_{plane} (r) =  i^\ell (2\ell +1 ) r j_\ell (kr).
\ee
To isolate the incoming and outgoing part, we write it in terms of Hankel functions
\be
h_\ell = j_\ell +i y_\ell.
\ee
In fact, we define the modified Hankel functions
\be
U_\ell (x) = ix h_\ell (x) = x (-y_\ell (x) + i j_\ell (x) ), 
\ee
in terms of which the radial $\ell$-plane wave function is
\be
u_{plane} (r) = \frac{1}{k}\, i^\ell \, \frac{2\ell +1}{2i} \left( U_\ell (kr) - {\bar U}_\ell (kr) \right)
\label{plane}
\ee
In the limit $kr \gg 1$ the functions $U_\ell$ behave as
\be
U_\ell (kr) \to (-i)^\ell e^{ikr}, 
\label{asym}
\ee
so $U_\ell$ is the outgoing part and ${\bar U}_\ell$ is the incoming part. $U_\ell$ and
its conjugate ${\bar U}_\ell$ separately satisfy the free radial equation
\be
\frac{d^2 U_\ell (kr)}{dr^2} + k^2 U_\ell (kr) = \frac{\ell (\ell +1)}{r^2} U_\ell (kr). 
\ee
The first few $U_\ell$ are
\be
U_0 (x)  = e^{ix},~~ U_1 (x) = e^{ix} \left(\frac{1}{x} -i\right) ,~~
U_2 (x) = e^{ix} \left( \frac{3}{x^2}- \frac{3i}{x}-1\right) ~,~{\rm etc.}
\label{us}
\ee

Returning to the solution of our problem, the radial wavefunction for $r>a$ (where
the potential vanishes) will still be a superposition of $U_\ell$ and ${\bar U}_\ell$.
The incoming part, proportional to ${\bar U}_\ell$, must be the same as in the plane
wave (\ref{plane}), while the outgoing part $U_\ell$ will have a different coefficient,
due to the existence of the outgoing scattered wave. The extra scattering part will
have the asymptotic  form
\be
\psi_{sc} \sim f \frac{e^{ikr}}{r}  ~~{\rm so}~~ u_{sc} \sim f e^{ikr}, 
\ee
as $r \to \infty$. From the asymptotic behavior (\ref{asym}) of $U_\ell$ we see
that the extra scattering part must be of the form
\be
u_{sc} = (-i)^\ell f U_\ell (kr)
\ee
so the full wavefunction for $r>a$ is
\bear
u_{out} &=& \left(\frac{1}{k}\, i^\ell \, \frac{2\ell +1}{2i}  +(-i)^\ell f \right) U_\ell (kr) - 
\frac{1}{k}\, i^\ell \, \frac{2\ell +1}{2i} {\bar U}_\ell (kr) \cr
&=& \frac{1}{k}\, i^\ell \, \frac{2\ell +1}{2i} \left( b \, U_\ell (kr) - {\bar U}_\ell (kr) \right), 
\label{out}
\eear
where
\be
b = 1 + (-1)^\ell \frac{2ik}{2\ell +1} \, f.
\ee
From unitarity, $b$ must be a pure phase (since incoming and outgoing waves must have
equal amplitudes) and it defines the scattering phase shift, while the above equation
relates the scattering amplitude to the scattering phase shift (and leads, eventually,
to the forward-scattering formula for the cross-section).

Solving the full problem for the radial wavefunction amounts to matching the interior
solution (\ref{in}), defined in terms of the solution $u$ and $A$, to the exterior solution
(\ref{out}) at $r=a$. Equating the values of the wavefunctions as well as their derivatives
on either side of $r=a$ we obtain
\bear
B u(a) &=& b\, U_\ell (ka) - {\bar U}_\ell (ka), \cr
B u'(a) &=& k b\, U'_\ell (ka) - k {\bar U}'_\ell (ka), 
\eear
where we defined
\be
B = (-i)^\ell \frac{2ik}{2\ell +1} A, 
\ee
from which we obtain
\bear
B &=& k \, \frac{{\bar U}_\ell U'_\ell- {\bar U}'_\ell U_\ell}{U_\ell u' -k U'_\ell u}, \cr
&& \cr
b &=& \frac{{\bar U}_\ell u' - k {\bar U}'_\ell u}{U_\ell u' -k U'_\ell u}, 
\eear
where we suppressed the dependence on $a$ or $ka$ to alleviate the form. Since
$u$ is real, it is clear that $b$ is a pure phase since ${\bar b} = b^{-1}$. The scattering
amplitude $f$ and the coefficient $A$ are determined through their relation with $b$
and $B$.

As a check, we apply the formulae for $\ell=0$. In that case, $U_0 =e^{ikr}$
and $A=\psi(0)$. We obtain
\be
B = \frac{2ik}{u' (a) -ik u(a) }e^{-ika} ~\Rightarrow~
A= \psi (0) = \frac{e^{-ika}}{u' (a) -ik u(a) }, 
\label{A}
\ee
which is our earlier result, while
\be
b = e^{-2ika} \frac{u' (a) +ik u(a)}{u' (a) -ik u(a)}~\Rightarrow~
f =e^{-ika} \frac{1}{k} \frac{k \cos (ka) \, u(a) - \sin (ka) \, u' (a)}{u' (a) -ik u(a)}, 
\ee
In the limit $ka \to 0$ this gives
\be
f =\frac{u(a)}{u' (a) -ik u(a)}, 
\label{f}
\ee
as before.
 
We can now tackle the issue of whether higher angular momenta contribute to the
scattering amplitude for $a$ very small. We will work out the case $\ell =1$, the others
being qualitatively similar.

For $\ell =1$ we can substitute the explicit form of $U_1$ from (\ref{us}) and obtain for $b$
\be
b = e^{-2ika} \frac{(1+ika - k^2 a^2 ) u(a) + a(1+ika) u' (a)}
{(1-ika - k^2 a^2 ) u(a) + a(1-ika) u' (a)}. 
\ee
So the scattering phase shift is $-2ka$ plus twice the phase of the complex number in the numerator.
We observe that for $ka \ll 1$ this phase will be negligible (and in fact of order $k^3 a^3$)
{\it no matter what the scaling of $u(a)$ and $u' (a)$ with $a$}. So, completely generically,
the contribution of the $\ell =1$ sector to the scattering amplitude is vanishingly small.
The only possibility to get an appreciable phase shift and scattering amplitude is for the
real and imaginary parts of the numerator to be of comparable magnitude. For this to happen
we need
\be
a u' (a) = - (1 - k^2 a^2 + \lambda k^3 a^3 ) u(a), 
\ee
for some finite $\lambda$. Thus, not only do we need $a u' (a)$ to be close to
$- u(a)$, but we need it tuned to such a value to an accuracy of order $k^3 a^3$ (!).
This is a tremendous level of fine-tuning. Moreover, it is $k$-dependent.
So even if it were to hold for some $k$, it would stop holding as soon as $k$
moves away from that value by the tiniest amount.

For higher $\ell$, a similar pattern emerges: the dominant terms are such that the
scattering amplitude vanishes, and an increasingly accurate fine tuning is required to have
an ``accidental" scattering resonance that would happen for only one specific
value of $k$, if at all. The final result is that only the $\ell =0$ sector
contributes.

\end{document}